\documentstyle[aps,epsfig]{revtex}
\newcommand{\be}{\begin{eqnarray}}
\newcommand{\ee}{\end{eqnarray}}

\begin{document} 

\title{    Why is a nucleon bound ? \footnote{To appear in Festschrift
    for Gerry Brown's 70.}}
\author{ Edward Shuryak}
\address{ Physics Department, SUNY at Stony Brook, NY 11794}
\maketitle

\begin{abstract}

  In a style of popular article, we
 discuss models of hadronic structure and their relation with
    models of the QCD vacuum and lattice
    simulations. Borrowing two main characters from G.Gamow,
    Mr.Thompson and Professor, we make a travel in the QCD vacuum. 
 Instanton-generated
interaction between quarks appear to be major player, they alone create
quark condensate, constituent quark masses and their bound states with
properties very close to those observed. Direct removal of
perturbative and confining forces (possible on the lattice by
``cooling'')
result in very small modification of hadrons. 
\end{abstract}

\section{Various models of hadronic structure}
 
  It is by now firmly established that
 strongly interacting particles  are made of quarks, which do not
 exist individually due to color confinement.  
The question {\it 
why} quarks form hadrons with precisely the  properties observed in
experiments is far from being quantitatively answered.  
However recent development have significantly clarified the issue,
some models seem to be qualitatively wrong and some are confirmed:
the instanton-induced interaction between quarks has emerged as a
major player, and existing models of vacuum and hadronic structure
based on it claim accuracy at 10-20\% level.

\widetext
\begin{figure}
\begin{center}
\leavevmode
\epsfig{file=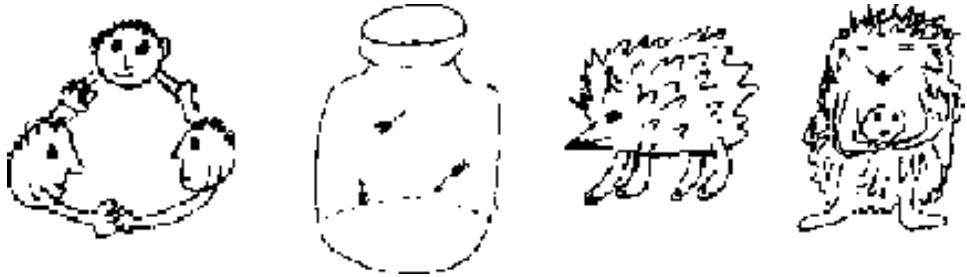, width=5in}
\end{center}
\caption{\label{hadronicmodels}
Four pictures of the nucleon structure: the non-relativistic quark model (a),
the MIT bag (b), the Skyrmion (c) and the ``chiral bag" (d).}
\end{figure}

  A brief introduction to old models of a nucleon structure
 is provided by Fig.\ref{hadronicmodels}.
  The  first picture (a) shows the essence of
the {\it nonrelativistic quark 
model} suggested in  60's. It shows a  family of three 
rather massive ``constituent quarks'' (with
 $M_{eff}= 300-400 MeV$) kept together by mutual
attraction (described by the potential V(r)). 

   Fig.\ref{hadronicmodels}(b) represents
 {\it the MIT bag} \cite{MITbag}, suggested in
the mid-70's, in the early days of QCD.
It is a completely different picture: 
the objects are nearly massless ``current'' quarks. They  
are not specifically attracted to each other, and perturbative
approach like Coulomb and magnetic spin-spin interactions are supposed to
be used inside the bag. The object exists
  because quarks are unable
to get  out of a ``bag'': they are  simply not admitted in
the ``physical vacuum" outside it.

   Fig.\ref{hadronicmodels} (c) corresponds to the so called {\it Skyrmion},  proposed 
in the early 60's \cite{Skyrme}
but becoming fashionable  in 80's, after  several  puzzling questions
have been
clarified.
 The ``hedgehog"  is made out of the pion 
field, with its pins representing  radially directed 
 isospin   $\vec \pi\sim \vec r$. 
If this objects rotates slowly (as it should,  provided the number of colors
$N_c>>1$
and baryons become parametrically heavy), it is the nucleon (spin and isospin
S=I=1/2), if more rapidly,
it becomes a $\Delta (S=I=3/2)$, etc. There are no quarks in this
picture at all.

   The last picture  Fig.\ref{hadronicmodels} (d) \cite{chiralbag} show a combination  of the
previous two:  it is a {\it chiral bag},
 surrounded by a hedgehog-shaped pion cloud. 
Gerry Brown and collaborators, who did the surgery, took care
and ensure that at the boundary the pressure and other important 
quantities like ``chiral current" are continuous, so the ``scar" is hardly 
physical. A smile of the hedgehog should remind us about the
so called ``Cheshire Cat Principle", according to which even the  location
of this ``scar" should be irrelevant. 

  Can all of those models be at least partially true at the same time?
It is very unlikely.  
  Looking at these models closer and trying to ignore differences in
language, one is still  puzzled 
by a completely different  physics involved.  
 For example,
according to the MIT bag
model, all hadronic properties directly follow from
confinement physics, with masses (and other dimensional quantities)
simply related to the bag constant B. The non-relativistic 
 potential model ascribe
  most of the hadronic mass to
the sum of
 ``constituent quark" masses, with only a small part coming from an
interaction. 
The Skyrmion picture 
 implies that quark 
and antiquark always  travel together, in a pion form: if 
 so, there is simply no place
for confinement left. 

   There are also many minor problems with those models, but one is a generic
 one, common to
   all of them. They do not follow the general wisdom which follows
from the solution of multiple quantum mechanics problems: one should
try to
understand  {\it the ground state} first, 
 then properties of the excitations will follow naturally. 
Hadrons we know most about (and therefore, most care about) 
like pions or nucleons are low-lying collective excitations of the QCD
vacuum,
like phonons in solids and nuclei. So, we have to focus our attention
on the underlying matter first.

    At this point it is fair to recall one model
which had actually followed this strategy closely. It 
is  the Nambu-Jona-Lasinio (NJL) model \cite{NJL} 
suggested in 1961, long before QCD and even before 
 quarks. It was 
inspired  by the BCS theory of superconductivity, and therefore
also  based on hypothetical
 attractive four-fermion interaction. It was shown that if
 it is
strong enough, it can rearrange the vacuum into a chirally 
asymmetric  (or ``superconducting") phase, with mesons analogous to
the Cooper 
pairs and (unconfined) quarks with reasonably large effective masses.

   The  main lesson we would like to discuss in this article is in
   fact the statement, that we now have convincing evidences that
 such kind of interaction
 actually  exists in QCD, and that its exact form and 
nature can be quantitatively obtained from a semiclassical theory
based
on instantons.

   Since we aim at readers which are not very familiar with
these methods and jargon used, we invite them 
to  a little travel through the QCD vacuum first, which will provide
some general picture.  
  Most of them  surely are familiar with Gamow's books,
explaining relativity and quantum mechanics to ``pedestrians"
with  unbeatable clarity\footnote{
They are also an excellent example of how much fun one 
may get in physics:  I was told by several 
colleagues that  these books  directly  
influenced their professional choice. Unfortunately, I was not among those
because Gamow's books  were not 
available in the part of the world where Gamow (and myself) come from.
  The explanations of  why this is the case, as well as a lively 
description of
some (failed) tunneling experiments he made himself, one can find 
in Gamow's autobiography ``My world line".}
. Let me borrow his style for this travel,
together with two main characters.

\section{Travelling  through the vacuum with Mr.Strange}

\widetext
\begin{figure}
\begin{center}
\leavevmode
\epsfig{file=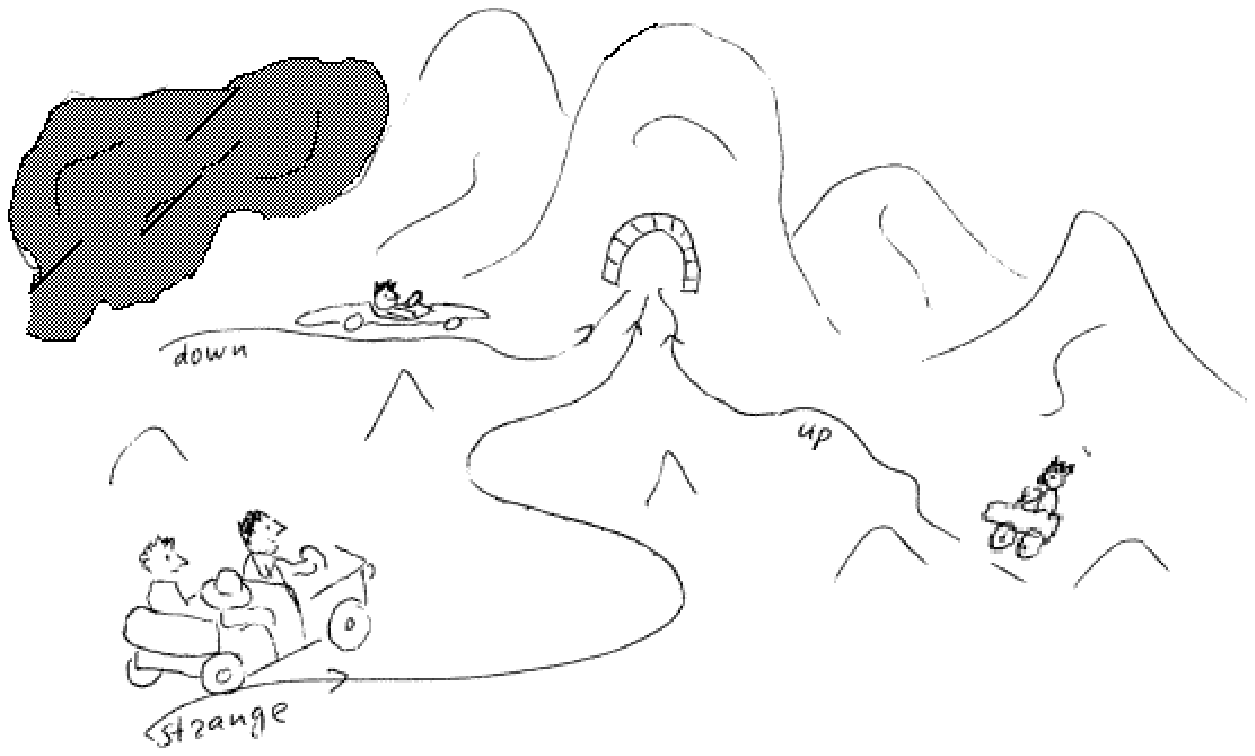, width=5in}
\end{center}
\caption{\label{travel}
Travel in the  QCD vacuum with Mr.Strange.
}
\end{figure}

    Mr.Thompson and Professor have found themselves   
 in a very colorful country, see Fig.\ref{travel}. They  were still 
looking around, when a car drove by.
A fat short fellow opened the door and said: ``I am Mr.Strange, and
 my job is to search this 
country. I was told you are interested in a tour, so please jump into the back 
seats".
 The car started, moving in an erratic unpredictable way,
avoiding larger bumps and jumping at many smaller ones. ``Do you see any 
road here?"-asked Mr.Thompson. ``We quarks do not need any,  just 
 take any path we like",    Mr.Strange
replied proudly. Puzzled by that, 
 Mr.Thompson thought for a moment and then
tried a simpler question: ``By the way, what color is your car?", but was again
 taken back by a strange cool 
reply:``Come on, this is not even a gauge invariant\footnote{Like coordinates in
general relativity, in QCD one is allowed to take any definition of 3 basic
quark charges, called colors, and change it at any point or moment. }
question!".

    Mr.Thompson gave up the questions and decided to look
around.  Other cars were travelling here and there, 
either jeeps, marked `Up',
or low sport models marked `Down'. After a while, a larger car appeared, 
with a nice     lady at the wheel, followed by a
jeep with a tiny little fellow in it. Mr.Strange made a signal and waved 
his hand to greet the lady before they disappeared behind a little hill.
 ``It is my first cousin Charm, 
 with that little fellow Anti-Up. He
spins around her all the time, but I do
not
trust him, though. Once I met him with another lady, known as 
Beauty, and he behaved in exactly the same way." 
Professor remarked,  with sudden
enthusiasm: "Oh, yes, this is what we call the Heavy Quark Symmetry\footnote{
Hydrogen atoms with a deuteron or triton instead of a proton have about
the same chemical properties. Similarly, with any heavy quark
(c,b or the recently discovered  t) one has about the same 
hadron. }".

\begin{figure}
\begin{center}
\leavevmode
\epsfig{file=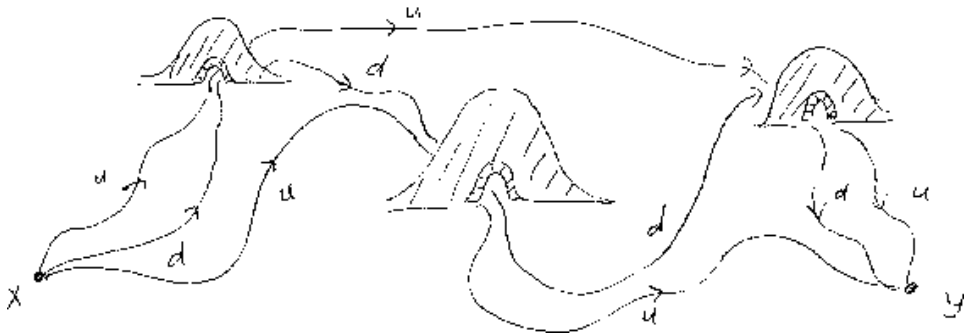, width=5in}
\end{center}
\caption{\label{travelindesert}
Travel in the ``cooled'' QCD vacuum, or the  
 ``desert" form.  
Examples  of typical paths used for 
proton
correlation function.
}
\end{figure}

   In a valley something like a race took place. A little crowd 
watched bunches of cars, each time consisting of
two ``ups" and a ``down" ones,
starting in regular intervals and disappearing in about the same direction
(see Fig.\ref{travelindesert}).
 ``It is the
measurements of the proton mass", - commented Mr.
Strange, ``they have done it for ages. A very dull job, 
I am glad I am not in the game\footnote{Mr.Strange hints that
studies of strange baryons, or hyperons, in which he could participate,
did not get much attention.}." 

\widetext

   Some object looking like a 
flat cloud but bent
in a complicated way
happened to be
close by. Professor became very agitated and asked whether it is 
possible to
drive through it.  
``Well, put your belts on - replied Mr.Strange,- its the 
storm". Indeed, the car was pulled in by a strong force,
 everybody's hair jumped up, 
lightnings was all around,  but in a second 
 it  was all over. ``This  cloud is known as a virtual string   
path", said Professor, ``They are as mysterious as the
ball lightnings. In spite of all the efforts,
nobody really understands what they are made of. Some say
 magnetic monopoles should be around, but I have not noticed any."
 
  A range of mountains blocked the way, but Mr.Strange did not slow down.
``I do not like mountain driving", said Mr.Thompson
still  frightened by the storm.
 ``We will take the tunnel, 
"- replied Mr.
Strange, `` Besides, we are not going to wait this 
time: see those fellows over there." He pointed toward
the tunnel entrance, where
two cars, ``up" and ``down" ones,
were moving in funny little circles. ``They are
 waiting for us. Due to the {\it First Tunneling Law}
 \cite{tHooft} no tunneling
is permitted unless
for a  complete set of quarks. By the way, I have seen 
 much higher mountains, made of W,Z rocks, and nobody was able to tunnel
through those ones
." ``I can tell you why," said Professor, ``according to the same Tunneling Law
one has to collect a company of 12 {\it weakly}
 interacting fermions, 
 with a representative
of all  quarks  and all 
 lepton families, electron, muon and tau. But even then, it is very
 unlikely
to happen \footnote{
The probability of tunneling in electroweak
 theory is very small, $exp(-{16\pi^2\over g^2_w})\sim 10^{-170}$,
where $g_w$ is the weak gauge coupling. It
 maybe happened once in the visible part of the Universe.
if it did, the baryon number was changed.
%It was hoped a few years ago that accelerating particles to tens-TeV
%energies  enhance this probability to an observable level, but
%further studies have indicated that the probability of this to happen 
%still remains about 60 orders of magnitude short!"
}.

   The tunneling itself took little time: a strong force pushed the 
car to another valley. But during it something very bizarre had
 happened: although Mr.Strange  continued driving,
he was sitting on the other side of the front seat!
 Mr.Thompson  looked at departing 
``up" and ``down" cars: the same thing has happened to them as well.
Mr.Strange noticed puzzled
 expression on his face in the mirror and  smiled: 
``Well, that is the {\it Second Tunneling Law}: anyone
who was right-handed  becomes left-handed,
 and vice versa.
Even  cars do that." Professor nodded and   made notes
in
his notebook (with a pen in his left hand, of course). He
said  he had studied this curious phenomenon, known as ``chiral
anomaly", for years going through multiple unclear papers,
% by 
%Schwinger, Adler, Bardeen 
%and Jakiw,
and how happy he is to see how the thing really works.
Mr.Thompson remarked: "Now I 
have an idea: how about putting suah a  tunnel between 
Britain and France?" 

   The rest of the journey was full of other adventures:
 it happen to be rather long. One of the reason
for that was the {\it Third Tunneling Law},  demanding that not a single
mountain should remain untunneled.
Finally the car returned to the spot
they had started from. Mr.Strange took a notebook and wrote down 
numbers from a device, which
 looked like a taxi  meter. ``My job is to evaluate
 how passable  this 
country is: it will appear in the next edition of the maps. Would you care 
for another trip?", said Mr.Strange.
  Mr.Thompson tried to escape, but 
Mr.Strange has said that it would be a very easy one,
 in the desert, and so they went along.

  And indeed, in was a completely different
 landscape (Fig.\ref{travelindesert}).
 Gone were all the large hills and little bumps: the
 country was basically a flat desert, with only a few mountains.
Those had also changed; they all now had the same rounded shape,
 looking like a set of domes.
%The weather was also fine: not a single 
%cloud was seen.
 The tunnels could be well seen from a distance,
and now, when they became used to them, Professor and Mr.Thompson
enjoyed them, as rides in Disneyland. Finally, after
Mr.Strange wrote down another set of
numbers, they thanked him once more and said goodbye to a  strange
colorful world.

\section{What can one  learn from this travel?}

 At this point, the reader is probably
 confused by details, or
even by the very goals of this travel. Well,
{\it some} explanations are coming.

 First of all, the ``landscapes" described above 
 represent the set of 
configurations of the ``colored" gauge field $A_\mu(x)$
 in 4-dimensional  space-time\footnote{Thus, it is more accurate to call 
 them {\it histories} of the field evolution.
The time running during the travel (such as
 shown
 by  Mr.Thompson's  watch) just  parameterizes
 the points on the quark path, and  in fact it is
unphysical. }. 
Furthermore, in order to simplify calculations one usually rotates 
 time into its imaginary axis, going into the so called Euclidean space-time
,
with the same metrics for all 4 coordinates: so there is 
no difference between them.
 Ensemble of ``landscapes" 
  represents the {\it wave function}
of the QCD ground state. Of course,  one should include
them  with the proper weight: 
$$ Weight= exp(-S_g(A)) \Pi_{q=u,d,s} det[i\gamma_\mu(\partial_\mu+igA_\mu)-m_q] 
$$
 
The first factor contains
(Euclidean) gauge field action  $S_g={1\over 2}\int d^4x (\vec E^2 + \vec B^2)$ 
 containing gluoelectric and gluomagnetic fields, while 
%\footnote{However, the
%field strength is expressed in terms of $A_\mu(x)$
% with the additional
% commutator term: since there are 3 charges rather than 1, all gluonic fields
%are actually $3\times 3$ matrices.}.
  the second factor is a product of
very complicated quantities,  the
``fermionic determinants", one for each kind of light quarks.
 It appears because we have chosen
 to integrate away all
fermionic degrees of freedom.
%We do not try to explain here  {\it how} they are calculated,
%but only note that
%those determinants  depend on the {\it whole picture} of the
%gauge field. (It is also worth mentioning that
% most of the money spent nowadays on ``lattice QCD" 
%goes into its evaluation.)
  The aim of
 Mr.Strange's travels is precisely
the evaluation of one of
those determinants. For example,
if he dislikes a particular field configuration, he can simply {\it veto}
it by giving it the zero value: then the configuration  will  be dropped
from the ensemble. 
%(This happens, for example, if one of the mountains is too 
%far from the rest, so it becomes very difficult to implement the Third Law.)
 
\setcounter{footnote}{0}
  One may find it surprising, that
 during our travel with Mr.Strange no direct
interaction between different quarks was seen. However, it does
{\it not} contradict to hadronic models discussed in the beginning
 because the {\it averaging
over the gauge fields} has not been done yet. As all quarks (i) avoid the same 
``bumps", (ii) suffer the same ``storms", and
(iii) tunnel through the same ``mountains".
Their
common adventures take them on similar path, or take them 
closer together. In a more 
conventional language, when one integrate over gauge field $first$,
 these are described 
as (i) the Coulomb-type (ii) confinement-related and (iii) instanton-induced 
forces, respectively.

   Now, suppose a proper set of field histories is collected: how can we
connect it to   mesons and baryons? In the same 
way as we study
``elementary excitations" of any matter: by observing the  propagation
of small perturbations. Say, to study ``phonons", one person can speak 
and another  listen. Similarly, one can inject to the vacuum
 few quarks at one point
and extract  them back at another: in this way one gets the so called
``point-to-point
correlation function". If this set  of quarks forms  a
bound state\footnote{For example, the
``races" observed during the travel above, was done for u,u,d quarks,
a set with quantum numbers of the nucleon.} they would travel together,
and  the behavior of the 
correlation function will reflect it. Furthermore, one can extract 
masses, wave functions, form-factors and other parameters which can be 
directly compared to the experimental data.

\widetext
\begin{figure}
\begin{center}
\leavevmode
\epsfig{file=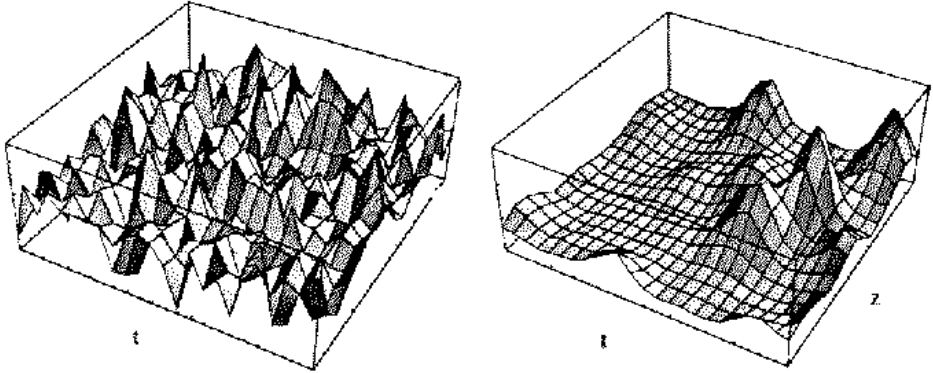, width=5in}
\end{center}
\caption{\label{cooling}
Sample of a  gauge configuration before (left) and after
  (right) cooling.In the latter case instantons are clearly seen.
 The quantity shown is the action density, and its scale (not shown) is
 two orders of magnitude larger on the left figure.
}
\end{figure}

\section{The instanton story}

  It is usual for  popular-style articles  to jump over
years of hard work of many people, proceeding directly to final conclusions.
Only few of them mention what was actually done, 
and even those  are usually related with only a couple of most recent
works. Alas, we have to follow the same
well trotted path,    with a brief 
 sketch of history.

    Lut us  start with conclusions. Above we have mentioned 3 types
    of
forces between quarks.
 Somewhat unexpectedly, quite different ones   bind
different quarks together,  
 depending mostly
on their {\it mass}.  Heavy b 
quarks make bound states, dominated by 
the Coulomb-type forces. 
 The ``charmed" $\bar c c$ pair forms mesons of the $J/\psi$ family,
using mainly the  confining potential.
 However, a nucleon (and other 
hadrons 
 made of light 
quarks) are mostly bound by 
forces induced by {\it tunneling}\footnote{The reader should also be warned
at this point, that this statement
 is  far from being the universally accepted.}.

   As usual, realization of that came  gradually,
 due to  a chain of seemingly
unrelated works. In 1975 Weinberg  \cite{U1} has pointed out that 
one particular meson, called
$\eta'$,  is about twice heavier than it should be due to
strange  quark masses\footnote{For example,
 its close relative $\eta$ meson contains
larger share of strangeness, but it is lighter.  By the way, $none$ of  
 models mentioned at the beginning 
can explain this phenomenon.}. At the same year, A.M.Polyakov with 
collaborators \cite{instanton} have found enigmatic solution of
Yang-Mills equations (QCD analog of Maxwell's ones).
Only later it was realized 
that it is a path describing the tunneling process. G. 't Hooft
\cite{tHooft} found the Tunneling Laws mentioned by Mr.Strange and
related it to ``anomalies''. Among other things, it was found 
%The first of them tells us, that instead of pair-wise interaction, we actually 
%have a {\it triple} one, including all u,d,s quarks into
%a complicated 6-fermion operator. That
that this interaction violates Weinberg's U(1) symmetry. Its effect
 in the $\eta'$ channel is {\it repulsive}, as needed, but in order to explain
the  puzzle in should be extremely strong\footnote{For example,
  counting powers of the number of colors one can find thatthe $\eta'$
  mass
is $O(N_c^{-1})$ and the nucleon one is $O(N_c)$. Naively for $N_c=3$ the former
should be an order of magnitude smaller, but experimentally both
masses are about the same. } Gradually it was realized, that if it is
that
strong in one channel, it cannot probably be unimportant in many
others as well. Although
 multiple attempts to derive properties of the
 instanton ensemble from first principles failed, but 
simple phenomenological model called the ``instanton liquid"\footnote{
It corresponds to a
``desert"  picture of vacuum fields, shown in Fig.\ref{travelindesert}}
has  simultaneously explained
large $\eta'$ mass, properties of the  pion and  
 few other vacuum 
parameters \cite{Shuryak_1982} .
 Its basic assumption was that, for whatever reason, in QCD vacuum
 there are many small-size instantons,
 $\rho\approx 1/3 fm$, with very strong field and
therefore semiclassical.

\begin{figure}
\begin{center}
\leavevmode
\epsfig{file=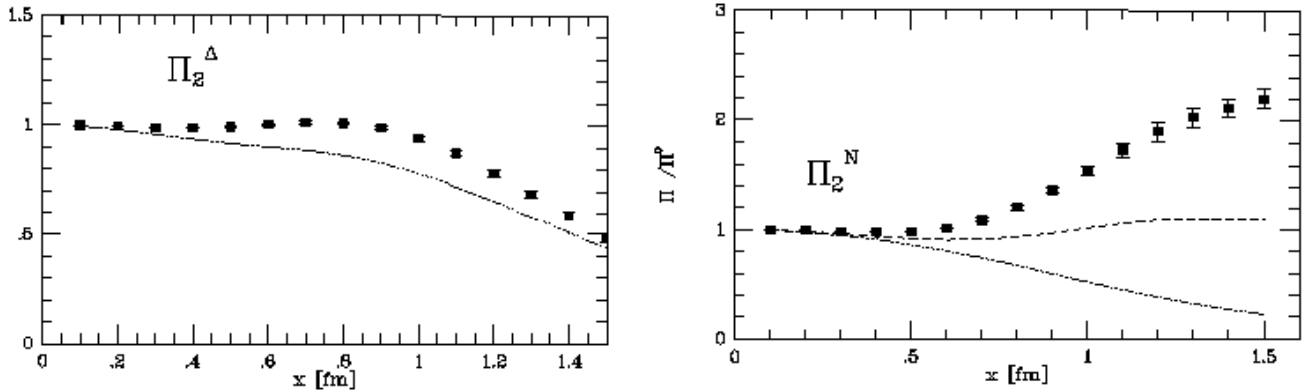, width=7in}
\end{center}
\caption{\label{baryons} Comparison of correlation functions for the
  nucleon and $\Delta$ channels: see text for explanation.}
\end{figure}

  Recently breakthrough is due to simultaneous
 attack on the problem by two teams, moving toward each other from opposite
  directions.
 The  Stony Brook group 
work out numerical methods capable to follow quark 
propagation (and correlators) in
a ``instanton liquid" up to rather large
distances $x\sim 1.5-2 fm$. Hadronic masses,
wave functions and other hadronic parameters \cite{cor} were calculated
in this model, with results  in  good 
agreement with data. 
%No free parameters were fixed compared to original paper
%a decade ago, but, for example, the pion mass was found to be 140 MeV and
Among other things, a nucleon was found to be deeply bound
state of constituent quarks, with the
right mass  $960\pm 30$  MeV. (And this is in the model $without$
confinement!)
Furthermore,  large 
 mass splitting between the nucleon and
 $\Delta$ isobar was found (in the model $without$ 
perturbative one-gluon exchange!). In Fig.\ref{baryons} we show the results
for the nucleon and $\Delta$ correlation function from \cite{cor}
(dots)
(lattice data which are not shown are in good agreement with them).  
Both are normalized in such a way that unit value correspond to free
propagation of massless quarks, and the argument $\tau$ is length of
the
quark travel in femtometers. An attrative interaction (the correlator rizes
above 1) is clearly seen in the nucleon case, but is absent for
 $\Delta$. The reason is (spin-0) u-d quark pair can tunnel together,
and thus are strongly attracted to each other: such pairs exist in the
nucleon but are absent in  $\Delta$.
For comparison, the dotted line show independent motion of three
constituent quarks, and the dashed on an independent motion of quark
and
bound scalar u-d diquark. None of those is close to the data points, which
can be well fitted by existence of the bound states.

\widetext

 The  MIT team \cite{Negele_etal}  started with the
{\it complete vacuum}, generated in lattice computer simulations,
and moved toward instanton physics gradually. As a first step,
the correlation function were measured: those were fund to be in
stunning agreement with instanton-based calculations. The second step 
\cite{CGHN_94} 
was application of
the so called ``cooling" algorithm, which makes lattice configuration smooth.
 Basically, only classical instanton field remained (see Fig.\ref{cooling}), while
 perturbative and confining forces were strongly suppressed.

% Fig.3 show how the field action is changed 
%during ``cooling": the picture changes from the one similar to Fig.2(a) to
%the ``desert", Fig.2(b), with a decrease in the action by more than
% two orders of 
%magnitude!
 Their first major  result is that
 all parameters of the ``instanton liquid" are reproduced, 
literally inside
the error
bars. Moreover, even the size distribution was recently measured
\cite{MS_94}, and it happens to be peaked at $\rho\approx 1/3 fm$.

But even more striking was an
observation that  detailed behavior
of the correlation functions measured after cooling have not
significantly
changed. In other words, the (lightest) hadrons 
{\it have survived  ``cooling"}! It has explicitly demonstrated, that by
sacrificing Coulomb and   confinement forces, one
still can get about
 the same hadrons\footnote{ 
About the largest deviation, at 20\% level,
is observed in
the shape and size
of the  pion wave function (Bethe-Salpeter 
amplitude), in which perturbative effects produced a characteristic
cusp
at small distances and confinement produces some extra suppression at
large ones.}.

  Let us now come back to  models of hadronic structure mentioned at
the beginning of the paper and
 try to connect them with these new results.
We have already commented in the Introduction
that NJL model is qualitatively correct,
 although  its original Lagrangian should be substituted 
by (much more complicated and non-local) instanton-induced Lagrangian
derived by 't Hooft. 
%(Certainly, the specific form of that 
%interaction was impossible to guess
%more than 30 years ago!}.
The similarity in fact goes even further:
 the BCS model is so successful 
because the  range of the interaction in
superconductors 
 is indeed much  smaller
 than the size of the Cooper pair. 
Similarly,  QCD instantons are
 relatively small compared to
sizes of most hadrons,  therefore
 this effective interaction can also be approximated 
by a local one.  Furthermore,
results for mesons can be approximately reproduced
by summing the same ``fish-type'' diagrams
(or solving Bethe-Salpeter equation) \cite{DP_86}.
Unfortunately, it is impossible to sum all of them
analytically, so the major tool remain computer simulations.

  These finding strongly contradict to some hadronic models mentioned
  above. The strongest case is against the MIT bag model: neither
  hadrons
are ``empty inside'' (true non-perturbative vacuum energy density is huge
compared to the MIT bag
constant),
nor  are quarks light or bound mainly by confinement.  
Even spin splittings seem to be not due to dipole spin-spin interaction!
   
 The chiral bag model,  with its large 
``pion cloud" around a small ``quark core", remains a reasonable picture 
(provided
the core is not  ``empty" but rather
 the place where non-perturbative 
fields are the strongest). 

 Of course, the story of hadronic 
structure is still far from being finished:
understanding and quantitative incorporation of confinement remains
the
long-standing challenging problem. 
Meanwhile
experiments   generate          new 
puzzles: ``spin crisis'' in the polarized nucleon,  
 large and polarized ``strange sea", strong
 isospin asymmetry of 
the ``u,d sea", etc. There are  hints that instantons  may  explain 
these puzzles as well, but this remains to be calculated.

\section{Weighting and melting the vacuum}

 In this last section let us consider
an important question about the QCD  {\it ground state  energy} of QCD.
As we will see, it is not so philosophical question as it sounds, but
rather a practical one.

  Still, let us start with a historical perspective. 
  An ancient philosopher would say, that 
the weight of an {\it empty}
bottle is  nothing else but the weight of the bottle itself.
A 16-th century physicist would be more careful:
he would point out the difference between an open
  bottle, with air, and the ``truly empty" one, with air pumped out of  it. A
20-th century theorist would comment that such ``truly empty" bottle
still contains zero-point
 perturbative fluctuations of all fields. In QED, after
infinities are subtracted, the so called
Casimir energy\footnote{ Since it  depends
on the size of the bottle and what it is made of, it can hardly be
ascribed to the QED vacuum itself, though.} remains.
In QCD this subtraction leads to a $finite$  difference  
 $\Delta\epsilon= \epsilon_{physical} - \epsilon_{perturbative}$.
By analogy say to a superconductor, which 
has lower energy compared to a normal metal, 
one expects that the physical vacuum 
 has {\it lower} energy than the perturbative one,
 $\Delta\epsilon<0$. 

  Theoretical expression, known as trace anomaly, relates
  $\Delta\epsilon$
to the so called gluon condensate.  
  In spite of intensive 
lattice simulations, we still know this quantity only very approximately.
However,  instantons alone produce a
{\it surprisingly
large} energy density,  $\Delta\epsilon\approx -1 GeV/fm^3$. 
It is about 20 times larger than the MIT bag constant value, and 
about 6 times larger than the mass density of nuclear matter!  
\setcounter{footnote}{0}
  Can one measure this vacuum energy in experiment? 
As we do not know how one can
``pump the non-perturbative fields out", we can at least
pump in a comparable amount of energy {\it into} some volume
and see what happens. It is predicted that the QCD vacuum
becomes the so called {\it quark-gluon plasma} at $T>T_c\sim 150 MeV$. 
At higher T, its energy density in believed to be
$\epsilon_{QGP}\sim T^4$, same as  the
``black body radiation"  (modulo a different number
 of degrees of freedom). This energy is counted from the 
  {\it perturbative} vacuum (the non-perturbative phenomena are
believed to be suppressed at high T), so comparing the two one may
find out  the ground state energy of QCD.
 
Because this energy density is so large,
 in order to ``melt the vacuum" and produce the new phase, 
one needs high energy colliders of heavy ions, such as the
Relativistic Heavy Ion Collider (RHIC), now under construction in
Brookhaven National Laboratory, or even the Large Hadron Collider 
to be built at CERN, Switzerland. 
These experiments  look also how ``melting" of hadronic states
takes place at the phase transition. If this is observed, it clearly sheds
some extra light at hadronic structure
as well. (For example, if the MIT bag model would be right,
hadrons would melt very easily,
at rather low energies such as Berkeley BEVALAC.) If the Skyrmion picture is 
correct, no trace of the nucleon above chiral restoration point
$T_c\approx 150 MeV$ is expected, where
the quark condensate and pion clouds are suppose to disappear.

    G.Brown
  and V.Koch \cite{KB_93} have analyzed
lattice data describing the QCD phase transition, they have concluded
that in fact only about half of the gluon condensate can ``melt''. 
How this other half (a ``hard glue'' or ``epoxy'', as Gerry Brown called it)
looks like? Why it does not create  quark condensate?
\widetext
\begin{figure}
\begin{center}
\leavevmode
\epsfig{file=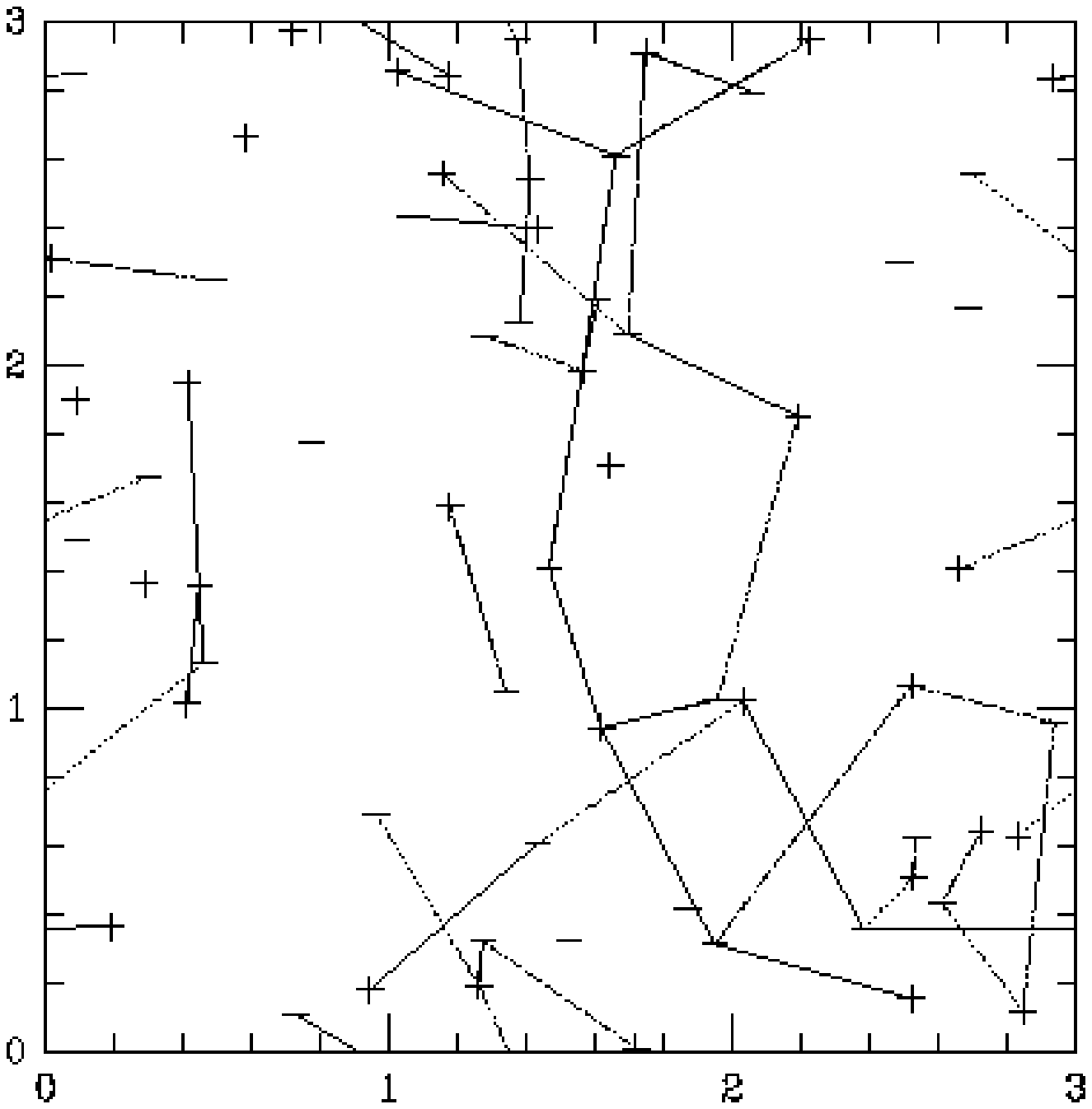, width=2.5in}
\end{center}
\end{figure}

\widetext
\begin{figure}
\begin{center}
\leavevmode
\epsfig{file=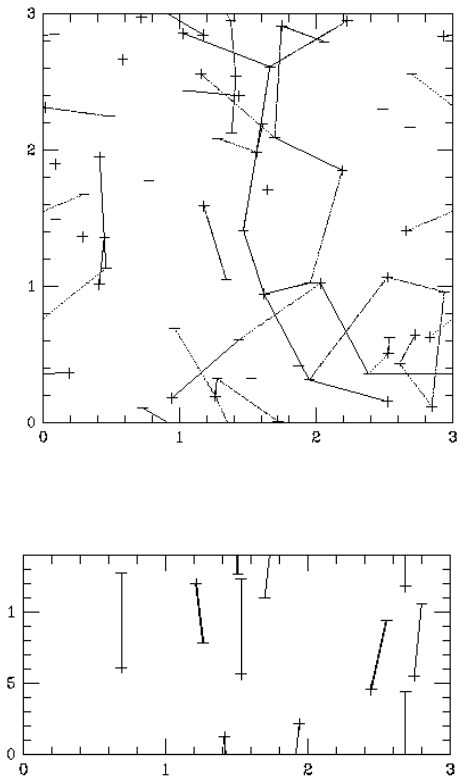, width=2.5in}
\end{center}
\caption{\label{molec}
Typical instanton configurations for $T=75$,  and
158 MeV. The plots show projections of a four dimensional
 $(3\Lambda^{-1})^3\times T^{-1}$ box into the 3-4 ($z$ axis-imaginary time) plane.
Instantons and antiinstanton positions are indicated by $+$ and $-$ symbols.
The lines correspond to strongest fermionic ``bonds''.}
\end{figure}

\begin{figure}
\begin{center}
\leavevmode
\epsfig{file=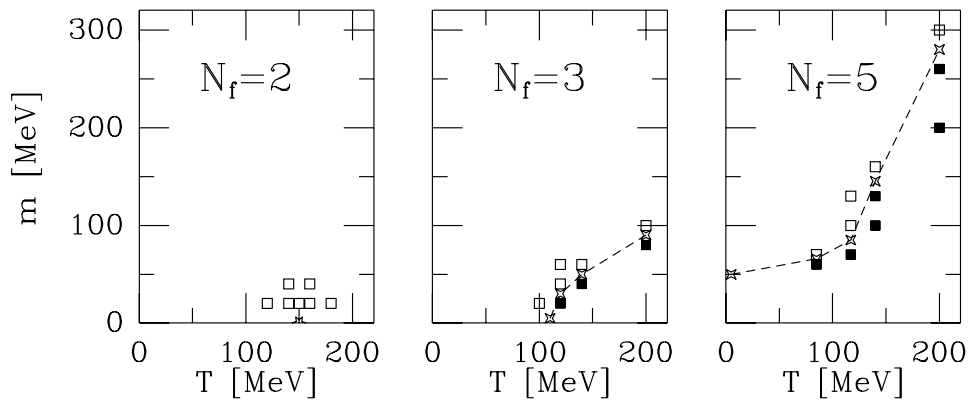, width=5in}
\end{center}
\caption{\label{phases}
Schematic phase diagram of the instanton liquid for
different numbers of quark flavors, $N_f$=2,3 and 5. We show the state
of chiral symmetry in the temperature-quark mass planes. In the figure
for $N_f$=2, open squares indicate points where we found large fluctuations
of the chiral condensate, the cross indicates the approximate location
of the singularity. In two other figures the open squares correspond
to non-zero
chiral condensate, while at solid it
is absent. The dashed lines connecting them
show the approximate location of the discontinuity line.
}
\end{figure}

The instanton-based theory makes quite specific predictions here as
well
(not yet directly tested on the lattice). It was found  \cite{molec}
that
at $T=T_c\approx 150 MeV$ relatively random instanton liquid undergo 
rapid  transition into a new phase, made of instanton-anti-instanton
molecules\footnote{Note a similarity to  Kosterlitz-Thouless
  transition in O(2) spin model in 2 dimensions: again one has paired
  topological objects, vortices, in one phase and random liquid in
  another.
The high and low-temperature phase exchange places, though.}
In a series of recent numerical simulations \cite{SS_95} 
it was found that this transition is indeed there, and its many
features and thermodynamics
 is consistent with available lattice data. In Fig.\ref{molec} we show a samble
 of configurations from this work, at different temperatures:
one can see how these molecules appear around critical temperature.
These molecules are the ``epoxy'', and the reason they do not create a
condensate is because they trap quarks inside them. Also,   
 they create  interaction
\cite{molec} between quarks (being even more similar 
to the original NJL one) even at $T>T_c$. This seem
to lead to existence
of some hadronic states (especially pions and its
chiral partner sigma)
surviving the phase transition! One more extension of those studies is
QCD
with larger number of flavors: when it exceed some critical value
 chiral symmetry is restored at T=0 (see Fig.\ref{phases}).  
Spectroscopy of this strange world with many flavors is
predicted to be entirely
dependent
on these molecules: it is exciting topic for future
lattice investigations.

\widetext

 Summarizing the main point of this last section:
 traditional studies of
 hadrons, as small perturbations of the QCD vacuum, can be
 supplemented by  experiments with the matter which is so
hot and dense, that it will have rather different (or even completely 
different, at $T>T_c$) excitations. Another way to perhapse similar
world, in which chiral symmetry is restored, is to add more quarks to
the QCD Lagrangian.

%3.The distribution of the action over the lattice in the complete (a) lattice
%configuration, and that after ``cooling" (b), from \cite{Negele_etal}.

%4.The pion and proton wave functions. The open and closed squares are the 
%results from the original and
%the ``cooled" lattice calculations, respectively.
%The points in the middle correspond to the instanton model. 

\end{document}